\documentclass[prl,longbibliography,showpacs,twocolumn,superscriptaddress,amscd,amssymb,verbatim]{revtex4-1}

\usepackage{graphicx}
\usepackage{textcomp}
\usepackage{lmodern}
\usepackage{color}
\definecolor{red}{rgb}{1,0,0}

\usepackage{epstopdf}

\newcommand{\CuD}{Zn-brochantite}
\newcommand{\CuH}{Zn-brochantite}

\begin{document}

\title{$\mu$SR insight into the impurity problem in quantum kagome antiferromagnets}
\author{M. Gomil\v sek}
\affiliation{Jo\v{z}ef Stefan Institute, Jamova c.~39, SI-1000 Ljubljana, Slovenia}
\author{M. Klanj\v sek}
\affiliation{Jo\v{z}ef Stefan Institute, Jamova c.~39, SI-1000 Ljubljana, Slovenia}
\author{M. Pregelj}
\affiliation{Jo\v{z}ef Stefan Institute, Jamova c.~39, SI-1000 Ljubljana, Slovenia}
\author{H. Luetkens}
\affiliation{Laboratory for Muon Spin Spectroscopy, Paul Scherrer Institute, CH-5232 Villigen PSI, Switzerland}
\author{Y. Li}
\affiliation{Department of Physics, Renmin University of China, Beijing 100872, People's Republic of China}
\author{Q. M. Zhang}
\affiliation{Department of Physics, Renmin University of China, Beijing 100872, People's Republic of China}
\author{A. Zorko}
\email{andrej.zorko@ijs.si}
\affiliation{Jo\v{z}ef Stefan Institute, Jamova c.~39, SI-1000 Ljubljana, Slovenia}

\date{\today}
\begin{abstract}
Impurities, which are unavoidable in real materials, may play an important role in the magnetism of frustrated spin systems with a spin-liquid ground state. 
We address the impurity issue in quantum kagome antiferromagnets by investigating ZnCu$_3$(OH)$_6$SO$_4$ (Zn-brochantite) by means of
muon spin spectroscopy.
We show that muons dominantly couple to impurities, originating from Cu-Zn intersite disorder, and that the impurity spins are highly correlated with the kagome spins, allowing us to probe the host kagome physics via a Kondo-like effect. 
The low-temperature plateau in the impurity susceptibility suggests that the kagome spin-liquid ground state is gapless. 
The corresponding spin fluctuations exhibit an unconventional spectral density and a non-trivial field dependence. 
\end{abstract}
\maketitle

The two-dimensional Heisenberg quantum kagome antiferromagnet (QKA), the paradigm of geometrical frustration, has been in the focus of attention for several years \cite{lacroix2011introduction,balents_spin_2010}. 
Theoretical studies have lately converged on a spin-liquid (SL) ground state \cite{ran2007projected,iqbal2011projected,yan_spin-liquid_2011,depenbrock_nature_2012,jiang_identifying_2012,iqbal2013gapless}, most likely with a finite gap to magnetic spinon excitations \cite{yan_spin-liquid_2011,depenbrock_nature_2012,jiang_identifying_2012}.
Although early experiments, on the contrary, spoke in favor of a gapless ground state \cite{olariu200817,faak2012kapellasite,clark2013gapless,li_gapless_2014}, experimental evidence of a gapped SL has also recently been presented \cite{fu2015evidence,han2015correlated}.
The confusion about the gap likely stems from the fact that all known QKA representatives contain a significant amount of defects that may affect their ground state, on top of other perturbations, such as magnetic anisotropy \cite{zorko2008dzyaloshinsky,cepas2008quantum,huh2010quantum,zorko2013dzyaloshinsky,he2015distinct} 
and exchange interactions beyond the nearest neighbors \cite{faak2012kapellasite,he2015distinct,domenge2005twelve,suttner2014renormalization,iqbal2015paramagnetism}. 
In particular, it is difficult to discern the magnetic response of the impurity spins from that of the kagome spins, which makes identification of their influence on the ground state elusive \cite{dommange2003static,rousochatzakis2009dzyaloshinskii,singh2010valence,poilblanc2010impurity,
kawamura2014quantum,shimokawa2015static}.

In herbertsmithite, the most extensively studied QKA representative to date \cite{mendels_quantum_2010}, a sizable amount (5-10\%) of Cu-Zn intersite disorder is present \cite{olariu200817,bert2007low, de2008magnetic}.
The general consensus is that these defects contribute significantly to the bulk magnetic response, but only at low energies ($E\lesssim 0.7$~meV) and at low temperatures ($T/J\lesssim 1/20$) \cite{han2015correlated,de2009scale,han2012fractionalized,nilsen2013low}.
Even though the defects are often described as quasi-free spin-1/2 impurities \cite{mendels_quantum_2010}, 
 their relation with the kagome spins is yet unsettled \cite{kermarrec2011spin,fu2015evidence,han2015correlated,shaginyan2016comment}.
It is important to resolve this issue, because a strong coupling would mean that defects could be intimately involved in the selection of a particular ground state of the QKA. 

We provide a unique perspective on the impurity problem by investigating another QKA representative, the recently synthesized ZnCu$_3$(OH)$_6$SO$_4$~(\CuH) \cite{li_gapless_2014}, which bears a resemblance to herbertsmithite in many respects. 
Despite a sizable average intraplane exchange interaction of $J=65$~K \cite{li_gapless_2014}, it remains magnetically disordered down to at least $T/J=1/3000$ \cite{Gomilsek2016instabilities}.
Moreover, just like herbertsmithite \cite{de2009scale}, it exhibits scale-free magnetic fluctuations at high $T$'s that are reminiscent of a critical correlated state \cite{Gomilsek2016instabilities}.
Finally, the two compounds are also similar regarding defects.
The amount of the Cu-Zn intersite disorder (6-9\%) and the effective Weiss temperature of impurities ($\sim$1~K) in \CuH~\cite{li_gapless_2014} both match those in herbertsmithite \cite{mendels_quantum_2010}.
However, the SL behavior of \CuH~is even more perplexing, as two SL instabilities have recently been found at different temperatures  \cite{Gomilsek2016instabilities}.
Here, the impurities may play an important role by pinning spinons at low $T$'s and thus enabling a spinon-instability mechanism.

We tackle the impurity problem in QKA's by performing muon spin relaxation and rotation ($\mu$SR) measurements on \CuD.
At low $T$'s, the muons are expected to be dominantly influenced by impurities through the long-range dipolar interaction \cite{Gomilsek2016instabilities}.
Such a scenario was indeed confirmed in herbertsmithite, where, however, the presence of a coupling between the impurity spins and the kagome spins remains unresolved \cite{kermarrec2011spin}.
If the impurities are coupled to the correlated state of the host, like they are in high-$T_{\rm c}$ superconductors and in many other strongly correlated electron systems \cite{alloul2009defects, georges2016beauty}, the muons may indirectly probe subtle correlations of the host state.
Moreover, the screening response of spinon excitations to impurities should strongly depend on the particular SL ground state \cite{kolezhuk2006theory},
which could then be determined by $\mu$SR. 
Here we demonstrate that in \CuD~muons detect the impurity magnetism at low $T$'s and that intrinsic correlations between the impurity spins and the kagome spins due to a Kondo-like effect are indeed present. 
The observed low-$T$ local-susceptibility plateau is consistent with a gapless host SL featuring a spinon Fermi surface \cite{kolezhuk2006theory}.
Moreover, this SL state is characterized by unusual spin dynamics with an intriguing magnetic-field dependence.
\begin{figure}[t]
\includegraphics[trim = 13mm 0mm 10mm 5mm, clip, width=1\linewidth]{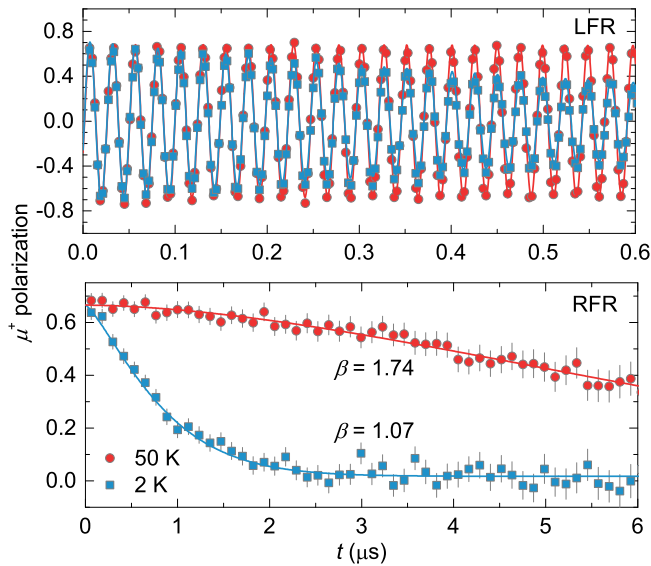}
\caption{The $\mu^+$ polarization in \CuD~in a TF of 0.3~T in the laboratory frame of reference (LFR; upper panel) and in the rotating frame of reference  (RFR; lower panel), the latter rotating with the Larmor frequency $\nu$ and thus displaying the envelope of the oscillations in the LFR. Symbols show the measurements and solid lines depict the fits, which are further explained in the text.
}
\label{fig1}
\end{figure}

$\mu$SR is a sensitive probe of magnetism that can easily distinguish between static and dynamic local magnetic fields $B_\mu$ \cite{yaouanc2011muon}.
In our previous $\mu$SR study in a small longitudinal field (LF) applied along the initial muon polarization, we showed that a dynamical magnetic state remains present in \CuD~at least down to 21~mK \cite{Gomilsek2016instabilities}.
Here, we build on this investigation by performing muon spin rotation measurements in a transverse magnetic field (TF) of 0.3~T, as well as muon spin relaxation measurement in various LF's.
The magnetic field $B_0$ was applied along the muon beam direction, with the initial muon polarization parallel to it in the LF experiment and tilted by $\sim$45$^\circ$ from it in the TF experiment.
$\mu$SR measurements were conducted on the General Purpose Surface Muon (GPS) and the Low Temperature Facility (LTF) instruments at the Paul Scherrer Institute, Switzerland on a $\sim$100\% deuterated \CuD~powder sample from the same batch as the one used in our previous study \cite{Gomilsek2016instabilities}.
A deuterated sample was used to minimize muon relaxation due to nuclear magnetic fields \cite{Gomilsek2016instabilities}.
The GPS instrument was operated in the veto mode, which minimized the amount of background signal below the detection limit, while the background in the LTF experiment ($\sim$20\% of the total signal) was determined by comparing the muon polarization curves from GPS and LTF and then subtracted from raw data.

\begin{figure}[b]
\includegraphics[trim = 9mm 18mm 3mm 5mm, clip, width=1\linewidth]{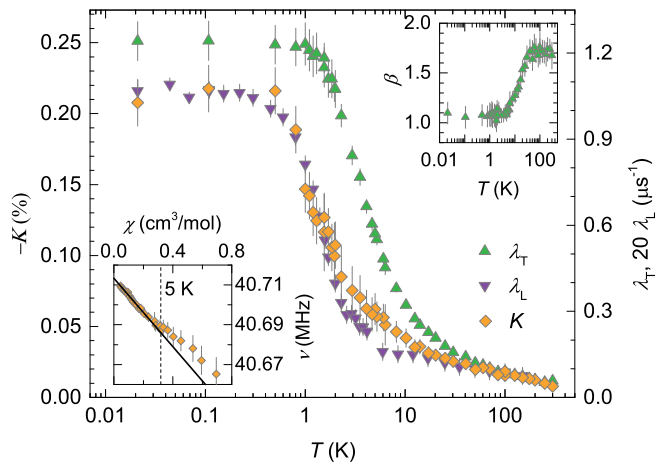}
\caption{The $T$-dependence of the Knight shift $K$ and the transverse relaxation rate $\lambda_{\rm T}$ in \CuD~in a TF of 0.3~T. 
The data for the longitudinal relaxation rate $\lambda_{\rm L}$ in a magnetic field of 8~mT are reproduced from Ref.~\onlinecite{Gomilsek2016instabilities}.
The lower inset displays the linear scaling of the mean precession frequency with the bulk susceptibility at low susceptibility values (high $T$'s). The extrapolation line yields $\nu_0=40.713$~MHz. 
The upper inset shows the $T$-dependence of the stretching exponent in the TF experiment.
}
\label{fig2}
\end{figure}

Typical TF $\mu$SR polarization curves for low and high $T$'s are shown in Fig.~\ref{fig1}.
In the TF experiment the muon polarization $P(t)$ precesses around the direction of $B_0$ with the Larmor frequency $\nu$.
The TF data are well described by the model
$P_{\rm TF}(t)=P_0\cos\left(2\pi\nu t -\phi \right){\rm e}^{-(\lambda_{\rm T}t)^\beta}$,
where $\phi\sim \pi/2$, $P_0=0.70$ is the projection of the initial muon polarization on the plane perpendicular to $B_0$, $\lambda_{\rm T}$ is the transverse muon spin relaxation rate and $\beta$ is the stretching exponent, which characterizes the distribution of static local fields $B_\mu$ induced by the applied field $B_0$.
In principle, $\lambda_{\rm T}$ is also affected by dynamic fluctuations of $B_\mu$. 
Yet, this effect is of the order of $\lambda_{\rm L}$, i.e., the relaxation that is purely dynamical in origin \citep{Gomilsek2016instabilities}. 
Since in \CuD~$\lambda_{\rm L}$ is more than an order of magnitude below $\lambda_{\rm T}$ (Fig.~\ref{fig2}), the latter is obviously dominated by static effects. 
     
We find that the shape of the $B_\mu$ distribution changes significantly between 40 and 4~K, as $\beta$ decreases progressively from its high-$T$ value of 1.75 to its low-$T$ value of 1.05 (upper inset in Fig.~\ref{fig2}).
This change is clearly seen in the bottom panel of Fig.~\ref{fig1}, which shows the polarization data in a rotating frame of reference.
The distribution of static $B_\mu$ is close to Gaussian ($\beta = 2$) at high $T$'s, but turns into Lorentzian ($\beta = 1$) at low $T$'s.
A Gaussian-like distribution of static fields is expected for dense disordered systems and a Lorentzian-like distribution for dilute systems \cite{yaouanc2011muon}.
The observed crossover is thus in line with the fact that in \CuD~the bulk magnetic susceptibility $\chi$ is dominated by dense kagome spins at high $T$'s, whereas at low $T$'s the diluted impurities yield the dominant contribution to $\chi$ \cite{li_gapless_2014}.
This suggests that the muons mainly detect the magnetism of the impurity spins at low $T$'s.

\begin{figure}[t]
\includegraphics[trim = 8mm 2mm 4mm 0mm, clip, width=1\linewidth]{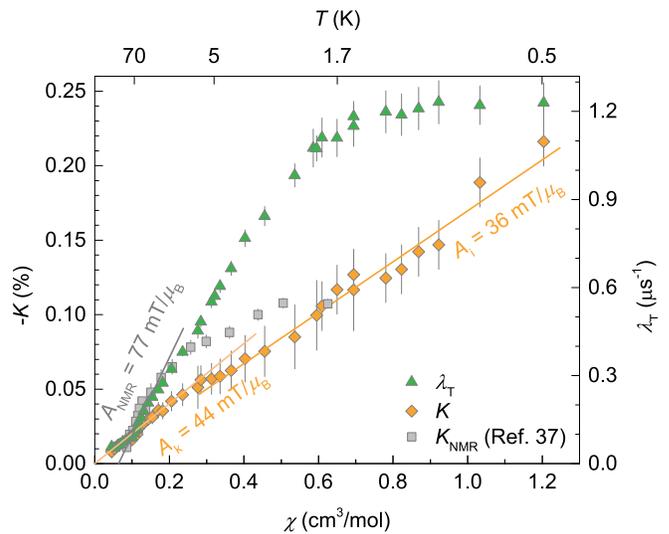}
\caption{The scaling of the Knight shift $K$ and the transverse muon relaxation rate $\lambda_{\rm T}$ with the bulk susceptibility $\chi$ in \CuD. 
The solid lines give the coupling $A_{\rm k}$($A_{\rm i}$) between the muons and the kagome (impurity) moments.
The NMR Knight-shift data $K_{\rm NMR}$ are reproduced from Ref.~\onlinecite{Gomilsek2016instabilities}.
}
\label{fig3}
\end{figure}

The precession frequency $\nu$ is shifted from $\nu_0=\gamma_\mu/2\pi\times B_0$ ($\gamma_\mu=2\pi\times 135.5\text{ MHz/T}$ is the muon gyromagnetic ratio) due to induced static $B_\mu$.
It scales linearly with the bulk susceptibility (lower inset in Fig.~\ref{fig2}) up to $\chi\sim 0.3$~cm$^3$/mol, i.e., down to $T\sim 5$~K, which  confirms the dominant coupling of the muons to the kagome spins at high $T$'s, in line with the high-$T$ value of $\beta$.
The corresponding Knight shift 
\begin{equation}
K=\frac{\nu-\nu_0}{\nu_0}=A\chi_\mu
\label{eq1}
\end{equation}
is proportional to the local susceptibility $\chi_\mu$, where $A$ is the  coupling constant between the electron and the muon magnetic moments. 
Even though $A$ is of dipolar origin, one expects $A\neq 0$ for powder Cu-based samples, because of a sizable anisotropy of the magnetic response  imposed by anisotropic $g$ factors, which typically span the interval 2.05--2.3 \cite{abragam1970electron}. 
The scaling of $K$ with $\chi$ changes around 5~K (Fig.~\ref{fig3}) and reveals that the average coupling of the muons to the impurity spins $A_{\rm i}=36$~mT/$\mu_{\rm B}$ is somewhat smaller than their coupling to the kagome spins $A_{\rm k}=44$~mT/$\mu_{\rm B}$.
The observed scaling of $K$ with $\chi$ below 5~K, where $\chi$ is mainly due to impurities \cite{li_gapless_2014}, and the change in $\beta$ unquestionably show that at these temperatures the $\mu$SR response is indeed dominated by impurities.
This is in sharp contrast to NMR measurements \cite{Gomilsek2016instabilities}, where the $^2$D Knight shift revealed that the magnetic coupling with the impurity spins is much smaller than with the kagome spins [note the low-$T$ leveling-off of the curve $K_{\rm NMR}(\chi)$ in Fig.~\ref{fig3}].
This can be explained by the difference in the dominant coupling mechanism of the two probes with electron magnetic moments, i.e., the short-ranged hyperfine coupling in NMR versus the long-ranged dipolar coupling in $\mu$SR.

$K$ saturates below $\sim$0.5~K, while the saturation of $\lambda_{\rm T}$ occurs at a $\sim$3 times higher temperature (Fig.~\ref{fig2}).
This dichotomy is even more obvious when comparing the variation of $\lambda_{\rm T}$ and $K$ with $\chi$ (Fig.~\ref{fig3}).
Both of these parameters reflect the static properties of the $B_{\mu}$ distribution; $K$ the average field value and $\lambda_{\rm T}$ the distribution width.
In a paramagnet with inhomogeneous broadening due to anisotropic powder averaging, one should find $\lambda_{\rm T}\propto K \propto \chi$.
However, if antiferromagnetic (AFM) correlations start developing they can affect the shape of the local field distribution.
Such an AFM correlated regime can therefore lead to a situation where $\lambda_{\rm T} \not\propto K \propto \chi$.
The saturation of $\lambda_{\rm T}$ below $\sim$1.5~K is thus a sign of quasi-static AFM correlations involving the impurity spins, as they dominate the system's magnetic response at low $T$'s.
This temperature corresponds well to the effective Weiss temperature  $\theta^{\rm CW}_{\rm i}\approx-1.2$~K of the impurity spins at low $T$'s \cite{li_gapless_2014}.
Furthermore, the $\lambda_{\rm T}(\chi)$ dependence exhibits an anomalous slope change already at 70~K (Fig.~\ref{fig3}), a temperature that is close to the Weiss temperature $\theta^{\rm CW}_{\rm k}=-79$~K of the kagome spins \cite{li_gapless_2014}.
This anomaly is thus likely due to the development of quasi-static spin correlations among the intrinsic kagome spins, which are also reflected in increased $^2$D NMR line width \cite{Gomilsek2016instabilities}.

The saturation of $K$ below $\sim$0.5~K (Fig.~\ref{fig2}), the lowest temperature of bulk $\chi$ measurements \cite{li_gapless_2014}, indicates that the local susceptibility in a magnetic field of 0.3~T reaches a plateau value of $\chi_\mu\sim 1.2$~cm$^3$/mol for $T\lesssim$0.5~K (Fig.~\ref{fig3}).
Comparing this with the calculated normalized saturation magnetization $M_{\rm s}/H=70$~cm$^3$/mol reveals an average magnetic moment of 0.017$\mu_{\rm B}$ per Cu site, which is much smaller than the full moment of $\sim$1$\mu_{\rm B}$.
It is significantly decreased from the full moment even if that is rescaled to 6--9\% spin-1/2 impurity concentration \cite{li_gapless_2014}.
Therefore, the impurity moments are obviously not saturated despite the plateau in $\chi_\mu$ below $\sim$0.5~K.
Furthermore, the plateau in $\lambda_L$, indicating a saturation in magnetic fluctuations even at 8 mT, occurs at the same temperature (Fig.~\ref{fig2}, Ref.~\onlinecite{Gomilsek2016instabilities}).
Both plateaus must thus be related to the low-$T$ SL state of the investigated compound which sets in around 0.6~K \cite{Gomilsek2016instabilities,li_gapless_2014}.
This unambiguously demonstrates a significant coupling of the impurity spins with the kagome spins.
Interestingly, a low-$T$ saturation of $\lambda_{\rm L}$ \cite{mendels2007quantum,kermarrec2011spin} and 
a small average magnetic moment \cite{ofer2009symmetry} were also found in herbertsmithite. 

\begin{figure}[t]
\includegraphics[trim = 0mm 18mm 12mm 5mm, clip, width=1\linewidth]{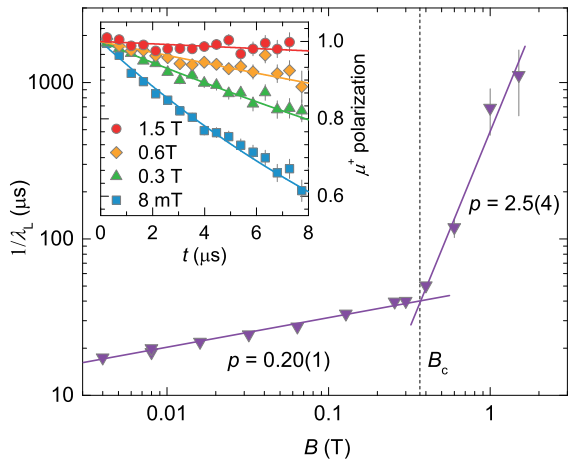}
\caption{The scaling of the inverse longitudinal muon relaxation rate with the applied LF in \CuD~at 110~mK. Two power-law regimes are indicated by solid lines. The inset shows the corresponding polarization curves (symbols) and stretched-exponential fits (solid lines).
}
\label{fig4}
\end{figure}

In order to corroborate the intimate connection of the impurity spins with the intrinsic kagome spins, as implied by the low-$T$ plateaus in $K(T)$ and $\lambda_{\rm L} (T)$, we performed additional LF $\mu$SR experiments at 110~mK, i.e., deep in the saturated regime.
The polarization curves (inset in Fig.~\ref{fig4}) are described by the stretched-exponential form $P_{\rm LF }(t)=P_0{\rm e}^{-(\lambda_{\rm L}t)^\beta}$, where the stretching exponent $\beta=0.86$ was fixed to the value found in our previous LF study \cite{Gomilsek2016instabilities}.
The inverse relaxation rate exhibits a power-law dependence on the applied field,
\begin{equation}
1/\lambda_{\rm L}\propto B^{p},
\label{eq3}
\end{equation}
with two notably different regions. 
The power $p=0.20(1)$ is found for magnetic fields below $B_c\sim 0.4$~T and $p=2.5(4)$ for higher magnetic fields.
We note that the muon relaxation due to nuclear magnetic fields in \CuD~is negligible for $B\gtrsim 4$~mT \cite{Gomilsek2016instabilities}, therefore, the observed relaxation is entirely due to magnetic fields arising from the electron magnetic moments.

The sublinear field dependence of $1/\lambda_{\rm L}$ for $B<B_c$ is rather unusual. 
The conventional exponential decay of the local-field autocorrelation function $\mathcal{S}(t)\propto e^{-t/\tau}$, characteristic of a single-correlation-time ($\tau$) Markovian local-field evolution, yields a Lorentzian spectral density $\mathcal{S}(\omega)$, where $1/\lambda_{\rm L}\propto\mathcal{S}(\omega)^{-1}\propto \left(\gamma_\mu B\right)^2$ \cite{hayano1979zero}.
The high-field value $p=2.5(4)$ is indeed close to 2.
The experimentally observed $p=0.2$ at low fields, however, suggests a more exotic spectral density $\mathcal{S}(\omega)\propto\omega^{-p}=\omega^{-0.2}$, where temporal correlations decay algebraically as $\mathcal{S}(t)\propto t^ {-(1-p)}=t^ {-0.8}$ \cite{pratt2006low}.
A similar situation with $p\leqslant 1$ was observed on a few occasions in different SL phases of spin chains \cite{pratt2006low}, pyrochlores \cite{keren2004dynamic}, and QKA's \cite{kermarrec2011spin, kermarrec2014spin}.

$\mathcal{S}(\omega)\propto\omega^{-0.2}$ could be due to a distribution of correlation times of the impurity spins.
The impurities could still be decoupled from the SL properties of the kagome spins and the power-law $\mathcal{S}(\omega)$ could be due to a presence of impurity-impurity couplings with a distribution of strengths arising from the randomness of  impurity positions. 
However, the drastic change of the spectral density at $B_c$ strongly suggests a more involved scenario.

The sudden sizable increase of $p$ at $B_c\sim 0.4$~T is rather intriguing.
It cannot be accounted for by field-induced polarization of the impurity spins, which would yield gradual changes.
It rather strengthens the conclusion from the saturation of the muon Knight shift that the impurities are coupled to the SL phase of the kagome spins. 
The properties of the SL phase obviously abruptly change at $B_c$.
Therefore, in addition to the well documented $T$-induced instabilities in \CuD~\cite{Gomilsek2016instabilities}, the low-$T$ SL state could also exhibit a field-induced instability at the critical field $B_c$.
This calls for future in-depth studies, as field-induced instability appears to be a common feature of frustrated antiferromagnets \cite{pratt2011magnetic,jeong2011field}, yet its origin remains unknown.

The correlations between the impurity and the kagome spins in the SL state can be understood as a Kondo-like effect, which can emerge on frustrated spin lattices from either fermionic or bosonic spinon excitations  \cite{kolezhuk2006theory,florens2006kondo,kim2008kondo,ribeiro2011magnetic}.
The resulting screening generally leads to Curie-Weiss behavior of impurity susceptibility at $T\gtrsim T_{\rm K}$, where the Kondo temperature $T_{\rm K}$ is of the order of $|\theta^{\rm CW}_{\rm i}|$  \cite{hewson1997kondo}.
Therefore, the observed low-$T$ Curie-Weiss behavior \cite{li_gapless_2014} does not require impurity-impurity interactions.
Below $T\sim T_{\rm K}$ the Curie-Weiss behavior should break down \cite{hewson1997kondo}, which nicely corresponds to the onset of the plateau in the impurity susceptibility $\chi_\mu$ at $\sim$0.5~K.  
Moreover, the Kondo effect can help to differentiate between different candidate SL ground states.
In particular, the finite $\chi_\mu$ at $T\rightarrow0$ observed in Zn-brochantite is consistent with predictions for a gapless SL with a spinon Fermi surface \cite{kolezhuk2006theory,ribeiro2011magnetic}, while it contradicts the divergent behavior expected for the gapless Dirac $U(1)$ SL's and the exponentially suppressed impurity response of gapped $Z_2$ SL's \cite{kolezhuk2006theory}.
Our observations are thus in accord with previous hints about the nature of the SL in Zn-brochantite from bulk heat-capacity measurements \cite{li_gapless_2014}.

The evidently gapless SL in Zn-brochantite is at odds with the apparently gapped SL in herbertsmithite \cite{fu2015evidence}.
Moreover, the experimentally detected correlations between the impurity and kagome spins in the SL state of Zn-brochantite stand in contrast to the recent inelastic neutron scattering results on herbertsmithite \cite{han2015correlated}, 
which advocated that a diluted impurity lattice with random-bond Heisenberg interactions may be effectively decoupled from the intrinsic kagome physics at $T/J\sim 1/100$.
The differences between the two compounds may have various origins; the perturbing magnetic anisotropies \cite{zorko2008dzyaloshinsky} are not necessarily the same, the position of the Zn (impurity) site is different \cite{li_gapless_2014} and the kagome lattice in Zn-brochantite is slightly distorted \cite{li_gapless_2014}. 

In conclusion, by performing $\mu$SR on \CuD~we have shown that this technique can comprehensively addresses the impurity magnetism that generally dominates the bulk response of QKA representatives at low $T$'s.
Our results reveal that the low-$T$ kagome-lattice SL state in the investigated compound is reflected in the magnetic behavior of the impurity spins, implying strong correlations between the impurity and the kagome spins.
Our findings thus suggest a Kondo-like description of impurity spins on the kagome lattice, a description that is also applicable to other quantum SL systems, e.g., the organic triangular-lattice antiferromagnets \cite{kolezhuk2006theory} and the Kitaev quantum SL's \cite{willans2010disorder,dhochak2010magnetic,sreejith2016vacancies}.
Based on theoretical predictions for the Kondo effect in SL's, details of the observed impurity behavior enable us to confirm a gapless SL with a spinon Fermi surface as the ground state of Zn-brochantite.
Impurities thus indeed prove to be salient local probes and may turn out to be essential for finally fully understanding the ground state of the QKA.

\acknowledgments{
The financial support of the Slovenian Research Agency (Program No.~P1-0125) and the Swiss National Science Foundation (SCOPES project IZ73Z0\_152734/1) is acknowledged.
The project has received funding from the European Union's Seventh Framework Programme for research, technological development and demonstration under the NMI3-II Grant number 283883.
Q.~M.~Z. was supported by the NSF of China and the Ministry of Science and Technology of China (973 projects: 2016YFA0300504).
We thank C.~Baines for his technical assistance at PSI.}

%

\end{document}